\documentclass
[prl,print£¬superscriptaddress,showpacs,onecolumn,showkeys]{revtex4}%
\usepackage{amsfonts}
\usepackage{amsmath}
\usepackage{amssymb}

\usepackage{fancyhdr}
\usepackage[raggedright]{titlesec}%
\providecommand{\U}[1]{\protect\rule{.1in}{.1in}}

\begin{document}
\preprint{HEP/123-qed}
\title{Measuring outcome correlation for Bell cat-state and geometric phase induced
spin parity effect }
\author{Yan Gu, Haifeng Zhang, Zhigang Song}
\affiliation{Institute of Theoretical Physics and Department of Physics, State Key
Laboratory of Quantum Optics and Quantum Optics Devices, Shanxi University,
Taiyuan, Shanxi 030006, China}
\author{J. -Q. Liang\footnote{*Corresponding author.}}
\affiliation{{Institute of Theoretical Physics and Department of Physics, State Key
Laboratory of Quantum Optics and Quantum Optics Devices, Shanxi University,
Taiyuan, Shanxi 030006, China} }
\affiliation{{*jqliang@sxu.edu.cn}}
\author{L. -F. Wei}
\affiliation{State Key Laboratory of Optoelectronic Materials and Technologies, School of
Physics and Engineering, Sun Yat-Sen University, Guangzhou 510275, China}
\affiliation{Quantum Optoelectronics Laboratory, School of Physics and Technology,
Southwest Jiaotong University, Chengdu 610031, China}

\begin{abstract}
In terms of quantum probability statistics the Bell inequality (BI) and its
violation are extended to spin-$s$ entangled Schr\"{o}dinger cat-state (called
the Bell cat-state) with both parallel and antiparallel spin-polarizations.
Except the spin-$1/2$ the BI is never ever violated by the Bell cat-states
with the measuring outcomes including entire Hilbert space. If, on the other
hand, measuring outcomes are restricted in the subspace of spin coherent state
(SCS), a universal Bell-type inequality (UBI), $p_{s}^{lc}\leq0$, is
formulated in terms of the local realistic model. We observe a spin parity
effect that the UBI can be violated only by the Bell cat-states of
half-integer but not the integer spins. The violation of UBI is seen to be a
direct result of non-trivial Berry phase between the SCSs of south- and
north-pole gauges for half-integer spin, while the geometric phase is trivial
for the integer spins. A maximum violation bound of UBI is found as
$p_{s}^{\max}$=1, which is valid for arbitrary half-integer spin-$s$ states.

\end{abstract}
\volumeyear{year}
\volumenumber{number}
\issuenumber{number}
\eid{identifier}
\date[Date text]{date}
\received[Received text]{date}

\revised[Revised text]{date}

\accepted[Accepted text]{date}

\published[Published text]{date}

\startpage{101}
\endpage{102}
\maketitle
\preprint{ }



\thispagestyle{fancy}

\section{1. Introduction}

Non-locality{\textsuperscript{\cite{non1,non2,non3}}} as one of the most
peculiar characteristic of quantum mechanics does not coexist with the
relativistic causality in our intuition of space and time. The two-particle
entangled state is a typical example of non-locality, which as a matter of
fact was originally considered by Einstein, Podolsky, and Rosen (EPR) to
question the completeness of quantum mechanics.{\textsuperscript{\cite{EPR}}}
Since it leads to apparently contradictory results with the locality and
reality criterion in classical theory.{\textsuperscript{\cite{EPR}}}
Nevertheless, the entanglement has become an essential ingredients in quantum
information and computation.{\textsuperscript{\cite{infor1,infor2,infor3}}}

With the spin version of EPR argument Bell formulated a quantitative test of
non-local correlations{\textsuperscript{\cite{Bell}}} known as Bell's
inequality (BI), which is derived in terms of classical statistics with
assumptions of local hidden-variables{\textsuperscript{\cite{hidden}}} and
measuring-outcome independence. The correlation properties of entangled states
are fundamentally different from the classical
world.{\textsuperscript{\cite{class1,class2,class3}}} The overwhelming
experimental evidence for the
violation{\textsuperscript{\cite{1,2,3,4,5,6,7,8}}} of BI opens up a most
intriguing aspect of non-locality in quantum mechanics, although the
underlying physics is obscure.{\textsuperscript{\cite{obscure}}} Following the
idea of Bell various extensions of the original BI were proposed such as
Clauser-Horne-Shimony-Holt{\textsuperscript{\cite{CHSH}}} (CHSH) and
Wigner{\textsuperscript{\cite{Wigner}}} inequalities. The experimental
tests{\textsuperscript{\cite{5,loop2,loop3}}} for the violation of BI have
been also reported using the spin entangled-states of a nitrogen-vacancy
defect in diamond.{\textsuperscript{\cite{diamond}}}

In our previous work{\textsuperscript{\cite{p1,p2}}} the original BI based on
the two-spin singlet is extended to a unified form, which is valid for the
general entangled states with both antiparallel and parallel polarizations.
The Bell-CHSH-Wigner inequalities and their violation are formulated in a
unified way by the spin coherent-state (SCS) quantum probability
statistics.{\textsuperscript{\cite{p3,p4}}} The density operator of entangled
state can be separated into the local and nonlocal parts. All inequalities are
derived in terms of the local part alone. While the nonlocal part leads to the
violation, which is a result of the coherent interference between two
components of the entangled state. The maximum violations are found for both
BI and Wigner inequality, which are shown to be equally convenient for the
experimental test.{\textsuperscript{\cite{p3,p4}}}

It has been pointed out that not all entangled states can be used to
demonstrate the Bell nonlocality.{\textsuperscript{\cite{Werner,uncover}}} A
natural question is wether or not the BI is violated by all entangled states?
By the explicit calculation it was shown that the BI is not violated by the
spin-$1$ entangled state.{\textsuperscript{\cite{p1,p2}}} We in the present
paper extend this investigation to entangled cat-state or Bell cat-state for
arbitrary spin-$s$. The cat state originated from
Schr\"{o}dinger{\textsuperscript{\cite{quantum theory}}} refers to any quantum
superposition of macroscopically distinct state called the macroscopic quantum
state, which gives rise to the minimum Heisenberg uncertainty relation. The
cat states received quite a lot of attention over the last
decade.{\textsuperscript{\cite{macro1,macro2}}} It is of fundamental
importance both theoretically and experimentally to examine the Bell
correlation in the entangled cat-state of arbitrary spin. The violation of BI
and phase effect of quantum states (Aharonov-Bohm phase, Berry phase) are both
considered as nonlocal phenomena in quantum mechanics. One goal of the present
study is tried to establish a relation between them in terms of the Bell cat-states.

In Section 2 the Bell correlation is examined for the spin-$3/2$ entangled
states, which do not lead to the violation of BI due to the cancellation of
non-local interference in the quantum probability statistics. The
non-violation of BI is approved for spin-$s$ entangled cat-states in Sec.3.
Sec.4 is devoted to spin parity effect with measurement restricted in the
subspace of SCSs only. A universal Bell-type inequality (UBI) is formulated
for the general entangled cat-states. In Section 5 we demonstrate the maximum
violation of UBI and the corresponding entangled states.

\section{2. Non-Violation of BI for Spin-3/2 Entangled State}

In the previous works{\textsuperscript{\cite{p1,p2}}} it was shown by the
explicit evaluation that the BI is not violated by the spin-$1$ entangled
state $\left\vert \psi\right\rangle =c_{1}|+1,-1\rangle+c_{2}|-1,+1\rangle$.
We extend the investigation to arbitrary spin-$s$ state. As an example we
first consider the spin-$3/2$ entangled state.

\subsection{\emph{2.1. Anti-parallel spin polarization}}

The spin-$3/2$ entangled state with anti-parallel polarization is%
\begin{equation}
\left\vert \psi\right\rangle =c_{1}|+\frac{3}{2},-\frac{3}{2}\rangle
+c_{2}|-\frac{3}{2},+\frac{3}{2}\rangle, \label{1}%
\end{equation}
in which the normalized superposition-coefficients are parameterized as
$c_{1}=e^{i\eta}\sin\xi$, $c_{2}=e^{-i\eta}\cos\xi$ with $\xi$, $\eta$ being
arbitrary real numbers. The measuring outcome-correlation-probability
respectively along two random directions $\mathbf{a}$ and $\mathbf{b}$ is
evaluated by the quantum probability statistics
\begin{equation}
P(a,b)=Tr[\hat{\Omega}(a,b)\hat{\rho}], \label{2}%
\end{equation}
in which%
\[
\hat{\Omega}(a,b)=(\hat{s}\cdot\mathbf{a})\otimes(\hat{s}\cdot\mathbf{b})
\]
is the two-spin correlation operator. The density operator $\hat{\rho
}=\left\vert \psi\right\rangle \langle\psi|$ of entangled state can be
separated into the local and non-local parts%
\[
\hat{\rho}=\hat{\rho}_{lc}+\hat{\rho}_{nlc},
\]
with%
\[
\hat{\rho}_{lc}=\sin^{2}\xi\left\vert +\frac{3}{2},-\frac{3}{2}\right\rangle
\left\langle +\frac{3}{2},-\frac{3}{2}\right\vert +\cos^{2}\xi\left\vert
-\frac{3}{2},+\frac{3}{2}\right\rangle \left\langle -\frac{3}{2},+\frac{3}%
{2}\right\vert ,
\]%
\[
\hat{\rho}_{nlc}=\sin\xi\cos\xi\left(  e^{2i\eta}\left\vert +\frac{3}%
{2},-\frac{3}{2}\right\rangle \left\langle -\frac{3}{2},+\frac{3}%
{2}\right\vert +e^{-2i\eta}\left\vert -\frac{3}{2},+\frac{3}{2}\right\rangle
\left\langle +\frac{3}{2},-\frac{3}{2}\right\vert \right)  .
\]
The measuring outcome correlation thus can be separated as the local
\begin{equation}
P_{lc}(a,b)=Tr[\hat{\Omega}(a,b)\hat{\rho}_{lc}], \label{lc-cr}%
\end{equation}
and nonlocal part
\begin{equation}
P_{nlc}(a,b)=Tr[\hat{\Omega}(a,b)\hat{\rho}_{nlc}]. \label{nlc-cr}%
\end{equation}
The complete eigenstate-set of projection spin-operators $\hat{s}%
\cdot\mathbf{a}$ and $\hat{s}\cdot\mathbf{b}$ must be taken into account in
the quantum probability statistics. The eigenstates of spin projection
operator along direction, say $\mathbf{r}$, are obtained by solving the
eigenstate equation, $\hat{s}\cdot\mathbf{r}\left\vert r_{m}\right\rangle
=m\left\vert r_{m}\right\rangle $ (in the unit convention $\hbar=1$) with the
magnetic quantum number $m=3/2,1/2,-1/2,-3/2.$ With the unit vector
parameterized by the polar and azimuthal angles\ $\theta_{r},\ \phi_{r}$ such
that $\mathbf{r}=(\sin\theta_{r}\cos\phi_{r},\sin\theta_{r}\sin\phi_{r}%
,\cos\theta_{r})$, we obtain the four eigenstates%
\begin{align*}
\left\vert r_{\frac{3}{2}}\right\rangle  &  =\cos^{3}\frac{\theta_{r}}%
{2}\left\vert +\frac{3}{2}\right\rangle +\sqrt{3}\sin\frac{\theta_{r}}{2}%
\cos^{2}\frac{\theta_{r}}{2}e^{i\phi_{r}}\left\vert +\frac{1}{2}\right\rangle
\\
&  +\sqrt{3}\sin^{2}\frac{\theta_{r}}{2}\cos\frac{\theta_{r}}{2}e^{2i\phi_{r}%
}\left\vert -\frac{1}{2}\right\rangle +\sin^{3}\frac{\theta_{r}}{2}%
e^{3i\phi_{r}}\left\vert -\frac{3}{2}\right\rangle ,
\end{align*}%
\begin{align*}
\left\vert r_{-\frac{3}{2}}\right\rangle  &  =\sin^{3}\frac{\theta_{r}}%
{2}\left\vert +\frac{3}{2}\right\rangle -\sqrt{3}\sin^{2}\frac{\theta_{r}}%
{2}\cos\frac{\theta_{r}}{2}e^{i\phi_{r}}\left\vert +\frac{1}{2}\right\rangle
\\
&  +\sqrt{3}\sin\frac{\theta_{r}}{2}\cos^{2}\frac{\theta_{r}}{2}e^{2i\phi_{r}%
}\left\vert -\frac{1}{2}\right\rangle -\cos^{3}\frac{\theta_{r}}{2}%
e^{3i\phi_{r}}\left\vert -\frac{3}{2}\right\rangle ,
\end{align*}%
\begin{align*}
\left\vert r_{\frac{1}{2}}\right\rangle  &  =\sqrt{3}\sin\frac{\theta_{r}}%
{2}\cos^{2}\frac{\theta_{r}}{2}\left\vert +\frac{3}{2}\right\rangle -\left(
1-3\sin^{2}\frac{\theta_{r}}{2}\right)  \cos\frac{\theta_{r}}{2}e^{i\phi_{r}%
}\left\vert +\frac{1}{2}\right\rangle \\
\text{ \ }  &  \ \ \ +\left(  1-3\cos^{2}\frac{\theta_{r}}{2}\right)
\sin\frac{\theta_{r}}{2}e^{2i\phi_{r}}\left\vert -\frac{1}{2}\right\rangle
-\sqrt{3}\sin^{2}\frac{\theta_{r}}{2}\cos\frac{\theta_{r}}{2}e^{3i\phi_{r}%
}\left\vert -\frac{3}{2}\right\rangle ,
\end{align*}%
\begin{align}
\left\vert r_{-\frac{1}{2}}\right\rangle  &  =\sqrt{3}\sin^{2}\frac{\theta
_{r}}{2}\cos\frac{\theta_{r}}{2}\left\vert +\frac{3}{2}\right\rangle +\left(
1-3\cos^{2}\frac{\theta_{r}}{2}\right)  \sin\frac{\theta_{r}}{2}e^{i\phi_{r}%
}\left\vert +\frac{1}{2}\right\rangle \nonumber\\
&  \text{ \ }+\left(  1-3\sin^{2}\frac{\theta_{r}}{2}\right)  \cos\frac
{\theta_{r}}{2}e^{2i\phi_{r}}\left\vert -\frac{1}{2}\right\rangle +\sqrt
{3}\sin\frac{\theta_{r}}{2}\cos^{2}\frac{\theta_{r}}{2}e^{3i\phi_{r}%
}\left\vert -\frac{3}{2}\right\rangle , \label{set}%
\end{align}
in which only the two states $\left\vert r_{\pm s}\right\rangle $\textbf{ (
}$s=3/2$ in the present case) are known as the
SCSs{\textsuperscript{\cite{p1,p2,p3,p4}}} (macroscopic quantum states)
satisfying the minimum uncertainty relation. For the measurement along two
random directions $\mathbf{a}$ and $\mathbf{b}$ the outcome correlation
probability Eq.(\ref{2}) is evaluated by the trace over 16 base vectors, which
may be grouped as%
\begin{align*}
\left\vert +,+\right\rangle  &  =\left\{  \left\vert a_{\frac{1}{2}}%
,b_{\frac{1}{2}}\right\rangle ,\left\vert a_{\frac{1}{2}},b_{\frac{3}{2}%
}\right\rangle ,\left\vert a_{\frac{3}{2}},b_{\frac{1}{2}}\right\rangle
,\left\vert a_{\frac{3}{2}},b_{\frac{3}{2}}\right\rangle \right\} \\
\left\vert +,-\right\rangle  &  =\left\{  \left\vert a_{\frac{1}{2}}%
,b_{-\frac{1}{2}}\right\rangle ,\left\vert a_{\frac{1}{2}},b_{-\frac{3}{2}%
}\right\rangle ,\left\vert a_{\frac{3}{2}},b_{-\frac{1}{2}}\right\rangle
,\left\vert a_{\frac{3}{2}},b_{-\frac{3}{2}}\right\rangle \right\} \\
\left\vert -,+\right\rangle  &  =\left\{  \left\vert a_{-\frac{1}{2}}%
,b_{\frac{1}{2}}\right\rangle ,\left\vert a_{-\frac{1}{2}},b_{\frac{3}{2}%
}\right\rangle ,\left\vert a_{-\frac{3}{2}},b_{\frac{1}{2}}\right\rangle
,\left\vert a_{-\frac{3}{2}},b_{\frac{3}{2}}\right\rangle \right\} \\
\left\vert -,-\right\rangle  &  =\left\{  \left\vert a_{-\frac{1}{2}%
},b_{-\frac{1}{2}}\right\rangle ,\left\vert a_{-\frac{1}{2}},b_{-\frac{3}{2}%
}\right\rangle ,\left\vert a_{-\frac{3}{2}},b_{-\frac{1}{2}}\right\rangle
,\left\vert a_{-\frac{3}{2}},b_{-\frac{3}{2}}\right\rangle \right\}  ,
\end{align*}
where the notation $\left\vert a_{m},b_{m^{\prime}}\right\rangle $ denotes
product eigenstates of spin projection operators respectively along directions
$\mathbf{a}$ and $\mathbf{b}$ with eigenvalues $m,m^{\prime}$. Each group
including four eigenstates corresponds to one measuring outcome state of the
spin-$1/2$ case. The measuring outcome correlation-probability of the local
part is derived from Eq.(\ref{lc-cr}) by the straightforward but tedious
algebra as
\begin{equation}
P_{lc}(a,b)=-\frac{9}{4}\cos\theta_{a}\cos\theta_{b}. \label{lc}%
\end{equation}
While the nonlocal part of correlation evaluated from Eq.(\ref{nlc-cr})
vanishes by the quantum probability statistics with the complete set of
eigenstates%
\[
P_{nlc}(a,b)=0.
\]
We may define a normalized correlation such that%
\begin{equation}
p(a,b)=\frac{P(a,b)}{s^{2}}. \label{nor}%
\end{equation}
Then the typical Bell correlation is recovered{\textsuperscript{\cite{p1,p2}}}%
\[
p(a,b)=p_{lc}(a,b)=-\cos\theta_{a}\cos\theta_{b}.
\]
The BI is not violated by the entangled cat-state of spin-$3/2$ with
antiparallel polarization.

\subsection{\emph{2.2. Parallel spin polarization}}

For the spin-$3/2$ entangled state with parallel polarization%
\[
\left\vert \psi\right\rangle =c_{1}|+\frac{3}{2},+\frac{3}{2}\rangle
+c_{2}|-\frac{3}{2},-\frac{3}{2}\rangle
\]
the local and non-local parts of state density operator become%
\[
\hat{\rho}_{lc}=\sin^{2}\xi\left\vert +\frac{3}{2},+\frac{3}{2}\right\rangle
\left\langle +\frac{3}{2},+\frac{3}{2}\right\vert +\cos^{2}\xi\left\vert
-\frac{3}{2},-\frac{3}{2}\right\rangle \left\langle -\frac{3}{2},-\frac{3}%
{2}\right\vert ,
\]%
\[
\hat{\rho}_{nlc}=\sin\xi\cos\xi\left(  e^{2i\eta}\left\vert +\frac{3}%
{2},+\frac{3}{2}\right\rangle \left\langle -\frac{3}{2},-\frac{3}%
{2}\right\vert +e^{-2i\eta}\left\vert -\frac{3}{2},-\frac{3}{2}\right\rangle
\left\langle +\frac{3}{2},+\frac{3}{2}\right\vert \right)  .
\]
The normalized local correlation probability is
\[
p_{lc}(a,b)=\cos\theta_{a}\cos\theta_{b},
\]
which is again exactly the same as Bell correlation of spin-$1/2$ entangled
state with parallel spin polarization.{\textsuperscript{\cite{p2,p4}}} The
nonlocal part vanishes by the quantum probability statistics $p_{nlc}(a,b)=0$.
We conclude that the modified{\textsuperscript{\cite{p2}}} and
extended{\textsuperscript{\cite{p4}}} BIs%
\begin{equation}
|p(a,b)-p(a,c)|-|p(b,c)|\leq1 \label{ext}%
\end{equation}
are not violated at all by the entangled spin-$3/2$ cat state for both
antiparallel and parallel spin polarizations.

\section{3. Non-Violation of BI for Spin-$s$ Entangled State by the Quantum
Probability Statistics}

We now turn to the entangled cat-state{\textsuperscript{\cite{p1}}} of
arbitrary spin-$s$ with anti-parallel polarization%
\begin{equation}
\left\vert \psi\right\rangle =c_{1}|+s,-s\rangle+c_{2}|-s,+s\rangle.
\label{anti}%
\end{equation}
\textbf{ }It is impossible to obtain all $(2s+1)$ analytic eigenstates
of\textbf{ }the projection spin-operator $\hat{s}\cdot\mathbf{r}$. We evaluate
the measuring outcome correlation by the trace over eigenstates of spin
operator $\hat{s}_{z}$($\hat{s}_{z}|m\rangle=m|m\rangle$) instead%
\begin{align*}
P\left(  a,b\right)   &  =%
{\displaystyle\sum\limits_{m_{1},m_{2}}}
\langle m_{1},m_{2}|\hat{\rho}\hat{\Omega}(a,b)|m_{1},m_{2}\rangle\\
&  =P_{lc}(a,b)+P_{nlc}(a,b),
\end{align*}
which can be worked out by straightforward calculation. The local part is the
same form%
\[
P_{lc}\left(  a,b\right)  =-s^{2}\cos\theta_{a}\cos\theta_{b}.
\]
While the nonlocal part again vanishes by the quantum average over
$(2s+1)^{2}$-dimension base vectors.

For the entangled state with parallel spin polarization%
\begin{equation}
\left\vert \psi\right\rangle =c_{1}|+s,+s\rangle+c_{2}|-s,-s\rangle,
\label{para}%
\end{equation}
the local part of correlation is%
\[
P_{lc}\left(  a,b\right)  =s^{2}\cos\theta_{a}\cos\theta_{b}.
\]
The nonlocal part vanishes by the quantum average. We have the normalized
correlation probability%
\[
p\left(  a,b\right)  =p_{lc}\left(  a,b\right)  ,
\]
for both cases. The extended BI is not violated by the Bell cat states except
$s=1/2$.

\section{4. Berry Phase Induced Spin Parity Effect and UBI}

We have demonstrated by explicit calculation that the BI is not violated by
the entangled cat-states of spin-$s$ ($s\neq1/2$) in the quantum probability
statistics, in which the nonlocal part of measuring outcome correlation
vanishes by the average over complete set of eigenstates. It is a interesting
problem if the measurements are restricted in the subspace of SCSs, namely
only the maximum spin values $\pm s$ are measured.

\subsection{\emph{4.1. Measuring outcome correlation in the subspace of SCSs}}

The SCSs for projection spin-operator\textbf{ }$\hat{s}\cdot\mathbf{r}$ in the
direction of unit vector $\mathbf{r}$ ($\mathbf{r=a,b}$) can be derived from
the eigenstate equations%
\[
\hat{s}\cdot\mathbf{r}\left\vert \pm\mathbf{r}\right\rangle =\pm s\left\vert
\pm\mathbf{r}\right\rangle .
\]
The explicit forms of SCSs in the Dicke-state representation are given
by{\textsuperscript{\cite{p1,p2,Dicke}}}%
\begin{align*}
\left\vert +\mathbf{r}\right\rangle  &  =%
{\textstyle\sum\limits_{m=-s}^{s}}
\left(
\begin{array}
[c]{l}%
2s\\
s+m
\end{array}
\right)  ^{\frac{1}{2}}K_{r}^{s+m}\Gamma_{r}^{s-m}\exp\left[  i\left(
s-m\right)  \phi_{r}\right]  \left\vert m\right\rangle \\
\left\vert -\mathbf{r}\right\rangle  &  =%
{\textstyle\sum\limits_{m=-s}^{s}}
\left(
\begin{array}
[c]{l}%
2s\\
s+m
\end{array}
\right)  ^{\frac{1}{2}}K_{r}^{s-m}\Gamma_{r}^{s+m}\exp\left[  i\left(
s-m\right)  \left(  \phi_{r}+\pi\right)  \right]  \left\vert m\right\rangle
\end{align*}
in which%
\[
K_{r}^{s\pm m}=\left(  \cos\frac{\theta_{r}}{2}\right)  ^{s\pm m}%
\]
and%
\[
\Gamma_{r}^{s\pm m}=\left(  \sin\frac{\theta_{r}}{2}\right)  ^{s\pm m}.
\]
\textbf{ }The two orthogonal states\ $|\pm\mathbf{r}\rangle$\ are known as
SCSs of north- and south- pole gauges, in which a phase factor $\exp
[i(s-m)\pi]$ difference between two gauges plays a key role in the spin parity
effect.\textbf{ }The eigenstates of projection spin-operators\ $\hat{s}%
\cdot\mathbf{a}$\ and $\hat{s}\cdot\mathbf{b}$\ form\ a measuring-outcome
independent base-vectors, if the measurements are restricted in the maximum
spin-values, $\pm s$.\textbf{ }The four base-vectors are labeled as%
\begin{equation}
\left\vert 1\right\rangle =\left\vert +\mathbf{a},+\mathbf{b}\right\rangle
,\left\vert 2\right\rangle =\left\vert +\mathbf{a},-\mathbf{b}\right\rangle
,\left\vert 3\right\rangle =\left\vert -\mathbf{a},+\mathbf{b}\right\rangle
,\left\vert 4\right\rangle =\left\vert -\mathbf{a},-\mathbf{b}\right\rangle ,
\label{base}%
\end{equation}
for the sake of simplicity. The measuring outcome
correlation{\textsuperscript{\cite{p1,p2,p3,p4}}} evaluated by the trace over
the subspace of SCSs can be also divided into local and non-local parts
\[
P(a,b)=Tr[\hat{\Omega}(a,b)\hat{\rho}]=P_{lc}(a,b)+P_{nlc}(a,b),
\]
with the normalized correlation probabilities Eq.(\ref{nor}) given by%
\[
p_{lc}(a,b)=\rho_{11}^{lc}-\rho_{22}^{lc}-\rho_{33}^{lc}+\rho_{44}^{lc},
\]
and%
\[
p_{nlc}(a,b)=\rho_{11}^{nlc}-\rho_{22}^{nlc}-\rho_{33}^{nlc}+\rho_{44}^{nlc}.
\]
Where
\[
\rho_{ii}=\langle i|\hat{\rho}|i\rangle=\rho_{ii}^{lc}+\rho_{ii}^{nlc}%
\]
($i=1,2,3,4$) denotes matrix elements of the density operator.\textbf{ }

\subsection{\emph{4.2. Spin parity effect for entangled state with
antiparallel spin-polarization}}

For the antiparallel spin-polarizations Eq.(\ref{anti}), the density-matrix
elements of local part are obtained as%
\[
\rho_{11}^{lc}=\sin^{2}\xi K_{a}^{4s}\Gamma_{b}^{4s}+\cos^{2}\xi\Gamma
_{a}^{4s}K_{b}^{4s},
\]%
\[
\rho_{22}^{lc}=\sin^{2}\xi K_{a}^{4s}K_{b}^{4s}+\cos^{2}\xi\Gamma_{a}%
^{4s}\Gamma_{b}^{4s},
\]%
\[
\rho_{33}^{lc}=\sin^{2}\xi\Gamma_{a}^{4s}\Gamma_{b}^{4s}+\cos^{2}\xi
K_{a}^{4s}K_{b}^{4s},
\]%
\[
\rho_{44}^{lc}=\sin^{2}\xi\Gamma_{a}^{4s}K_{b}^{4s}+\cos^{2}\xi K_{a}%
^{4s}\Gamma_{b}^{4s}.
\]
The non-local parts are
\begin{align}
\rho_{11}^{nlc}  &  =\rho_{44}^{nlc}\nonumber\\
&  =\sin2\xi K_{a}^{2s}\Gamma_{a}^{2s}K_{b}^{2s}\Gamma_{b}^{2s}\cos\left[
2s\left(  \phi_{a}-\phi_{b}\right)  +2\eta\right]  \label{nlc11}%
\end{align}
and%
\begin{equation}
\rho_{22}^{nlc}=\rho_{33}^{nlc}=\left(  -1\right)  ^{2s}\rho_{11}^{nlc}.
\label{22}%
\end{equation}
It may be worthwhile to remark that the density matrix elements of nonlocal
part for the same-direction measurement of two spins differ with that of
opposite directions by a phase factor $(-1)^{2s}=\exp(i2s\pi)$, which resulted
from the geometric phase between SCSs of the north- and south- pole gauges.
The normalized local correlation probability is found as%
\begin{equation}
p_{lc}(a,b)=-\left(  K_{a}^{4s}-\Gamma_{a}^{4s}\right)  \left(  K_{b}%
^{4s}-\Gamma_{b}^{4s}\right)  . \label{local}%
\end{equation}
While the nonlocal part of correlation is simply
\begin{equation}
p_{nlc}(a,b)=2\left[  1-\left(  -1\right)  ^{2s}\right]  \rho_{11}^{lc},
\label{nlc}%
\end{equation}
which vanishes for integer spin-$s$ but not for the half-integer spin. This
spin parity effect is a result of geometric phase.

\subsection{\emph{4.3. Parallel polarizations}}

For the state with parallel polarization Eq.(\ref{para}), the\ density-matrix
elements of local part are modified as%
\[
\rho_{11}^{lc}=\sin^{2}\xi K_{a}^{4s}K_{b}^{4s}+\cos^{2}\xi\Gamma_{a}%
^{4s}\Gamma_{b}^{4s}%
\]%
\[
\rho_{22}^{lc}=\sin^{2}\xi K_{a}^{4s}\Gamma_{b}^{4s}+\cos^{2}\xi\Gamma
_{a}^{4s}K_{b}^{4s}%
\]%
\[
\rho_{33}^{lc}=\sin^{2}\xi\Gamma_{a}^{4s}K_{b}^{4s}+\cos^{2}\xi K_{a}%
^{4s}\Gamma_{b}^{4s}%
\]%
\begin{equation}
\rho_{44}^{lc}=\sin^{2}\xi\Gamma_{a}^{4s}\Gamma_{b}^{4s}+\cos^{2}\xi
K_{a}^{4s}K_{b}^{4s}. \label{el}%
\end{equation}
The four matrix elements of nonlocal part have the same relation with the
antiparallel case that%
\begin{align}
\rho_{11}^{nlc}  &  =\rho_{44}^{nlc}\nonumber\\
&  =\sin2\xi K_{a}^{2s}\Gamma_{a}^{2s}K_{b}^{2s}\Gamma_{b}^{2s}\cos\left[
2s\left(  \phi_{a}+\phi_{b}\right)  +2\eta\right]  , \label{nl}%
\end{align}
and $\rho_{22}^{nlc}=\rho_{33}^{nlc}=\left(  -1\right)  ^{2s}\rho_{11}^{nlc}%
.$The local correlation probability%
\begin{equation}
p_{lc}(a,b)=\left(  K_{a}^{4s}-\Gamma_{a}^{4s}\right)  \left(  K_{b}%
^{4s}-\Gamma_{b}^{4s}\right)  , \label{local2}%
\end{equation}
possesses a positive sign different from the antiparallel case Eq.(\ref{local}%
). The nonlocal correlation probability is also the same as Eq.(\ref{nlc})
$p_{nlc}(a,b)=2\left[  1-\left(  -1\right)  ^{2s}\right]  \rho_{11}^{nlc}$,
here the matrix element $\rho_{11}^{nlc}$ Eq.(\ref{nl}) is slightly different
from antiparallel polarizations in Eq.(\ref{nlc11}). The spin parity effect
holds for the parallel spin-polarizations. Particularly for $s=1/2$ the
measuring outcome correlations reduce to the well known
results{\textsuperscript{\cite{p1,p2,p4}}}.

\subsection{\emph{4.4. UBI for entangled state of half-integer spin}}

The original BI, which requires the total measuring-probability to be one, is
not suitable for the incomplete measurement in the SCS subspace, where the
measured probability is less than one. Following Bell we also consider the
three-direction ($\mathbf{a}$, $\mathbf{b}$, $\mathbf{c}$)\textbf{
}measurements. The UBI is then formulated as%
\begin{equation}
\left\vert p_{lc}\left(  b,c\right)  \right\vert \geq p_{lc}\left(
a,b\right)  p_{lc}\left(  a,c\right)  , \label{nbi}%
\end{equation}
which is suitable for both antiparallel and parallel polarizations. Notice the
forms of local correlation $p_{lc}$ in Eqs.(\ref{local},\ref{local2}), the
validity of the UBI is obvious for the local realistic model. The UBI is also
derived from classical statistics with the hidden variable in Appendix. To
find the maximum violation of UBI Eq.(\ref{nbi}) we may define a quantity of
probability difference that
\begin{equation}
p_{s}^{lc}=p_{lc}\left(  a,b\right)  p_{lc}\left(  a,c\right)  -\left\vert
p_{lc}\left(  b,c\right)  \right\vert \leq0. \label{ps}%
\end{equation}
Any positive $p_{s}$ indicates the violation of UBI.

The quantum correlation probability of measuring outcomes along $\mathbf{a}$,
$\mathbf{b}$ reads
\begin{equation}
p(a,b)=\mp\left(  K_{a}^{4s}-\Gamma_{a}^{4s}\right)  \left(  K_{b}^{4s}%
-\Gamma_{b}^{4s}\right)  +4\sin2\xi K_{a}^{2s}\Gamma_{a}^{2s}K_{b}^{2s}%
\Gamma_{b}^{2s}\cos\left[  2s\left(  \phi_{a}\mp\phi_{b}\right)
+2\eta\right]  \label{qc}%
\end{equation}
respectively for antiparallel and parallel spin-polarizations. It reduces to
the well known two-state result{\textsuperscript{\cite{p1,p2,p4}}} when
$s=1/2$. The UBI can be violated by the quantum correlation Eq.(\ref{qc}) for
half-integer spin-$s$ entangled states.

\section{5. Maximum Violation of UBI}

\subsection{\emph{5.1. Spin-1/2 entangled state}}

First of all we examine the maximum violation of UBI Eq.(\ref{ps}) by the
spin-$1/2$ entangled states. For the antiparallel spin-polarization, the
quantum correlation probability Eq.(\ref{qc}) becomes the known
result{\textsuperscript{\cite{p1,p4}}}%
\begin{equation}
p\left(  a,b\right)  =-\cos\theta_{a}\cos\theta_{b}+\sin2\xi\sin\theta_{a}%
\sin\theta_{b}\cos\left(  \phi_{a}-\phi_{b}+2\eta\right)  . \label{qc1/2}%
\end{equation}
Since the polar angle is restricted by $0\leq\theta\leq\pi,$ it is easy to
verify that the UBI can be violated by the quantum correlation probability
Eq.(\ref{qc1/2}). The quantum correlation-probability difference is bounded by
a maximum violation value $P_{s}^{\max}$ that%
\begin{align*}
p_{\frac{1}{2}}  &  =p\left(  a,b\right)  p\left(  a,c\right)  -\left\vert
p\left(  b,c\right)  \right\vert \\
&  \leq\cos\left(  \theta_{a}+\theta_{b}\right)  \cos\left(  \theta_{a}%
+\theta_{c}\right) \\
&  \leq1=P_{\frac{1}{2}}^{\max}.
\end{align*}
As a matter of fact when $\theta_{a}=\theta_{b}=\theta_{c}=\pi/2$, the
probability difference is%
\[
p_{\frac{1}{2}}=\left(  \sin2\xi\cos\left(  \phi_{a}-\phi_{b}+2\eta\right)
\right)  \left(  \sin2\xi\cos\left(  \phi_{a}-\phi_{c}+2\eta\right)  \right)
-\left\vert \sin2\xi\cos\left(  \phi_{b}-\phi_{c}+2\eta\right)  \right\vert .
\]
Choosing the entangled state angles $\xi=\eta=\left(  \pi/4\right)
\operatorname{mod}2\pi$, we have%
\[
p_{\frac{1}{2}}=\sin\left(  \phi_{a}-\phi_{b}\right)  \sin\left(  \phi
_{a}-\phi_{c}\right)  -\left\vert \sin\left(  \phi_{b}-\phi_{c}\right)
\right\vert .
\]
The maximum violation $P_{\frac{1}{2}}^{\max}=1$ is then realized with
measuring direction angles $\phi_{a}=\pi/2,\phi_{b}=\phi_{c}=0$.

The entangled state to realize the maximum violation is%
\begin{equation}
\left\vert \psi\right\rangle =\frac{1}{\sqrt{2}}\left(  e^{i\frac{\pi}{4}%
}|+\frac{1}{2},-\frac{1}{2}\rangle+e^{-i\frac{\pi}{4}}|-\frac{1}{2},+\frac
{1}{2}\rangle\right)  . \label{ap}%
\end{equation}
The three-direction measurements should be set up respectively with the polar
and azimuthal angles $\theta_{a}=\theta_{b}=\theta_{c}=\pi/2,$ $\phi_{a}%
=\pi/2,$ $\phi_{b}=\phi_{c}=0$, namely $\mathbf{a},\mathbf{b},\mathbf{c}$ are
perpendicular to the original spin polarization with $\mathbf{a}$ along
$y$-direction $\mathbf{b,c}$ along $x$-direction.

With the same analyses the state with parallel spin-polarization to realize
the maximum violation of UBI is seen to be%
\begin{equation}
\left\vert \psi\right\rangle =\frac{1}{\sqrt{2}}\left(  e^{i\frac{\pi}{4}%
}|+\frac{1}{2},+\frac{1}{2}\rangle+e^{-i\frac{\pi}{4}}|-\frac{1}{2},-\frac
{1}{2}\rangle\right)  , \label{p}%
\end{equation}
for which the three-direction measuring angles are the same as that of
antiparallel polarizations.

\subsection{\emph{5.2. Arbitrary half-integer spin-}$s$}

Including the nonlocal part the quantum correlation probability difference
defined in Eq.(\ref{qc}) is%
\begin{equation}
p_{s}=p\left(  a,b\right)  p\left(  a,c\right)  -\left\vert p\left(
b,c\right)  \right\vert , \label{bd}%
\end{equation}
any positive value of which indicates the violation of UBI. We now find the
maximum violation value $p_{s}^{\max}$, the corresponding state parameters
$\xi,\eta$, and the three measuring directions. From the spin-$1/2$ case we
know that the three measuring directions $\mathbf{a},\mathbf{b},\mathbf{c}$
are perpendicular to the original spin polarization, namely\ $\theta
_{a}=\theta_{b}=\theta_{c}=\pi/2$, which leads to%
\[
K_{r}^{4s}-\Gamma_{r}^{4s}=0,
\]
with $\mathbf{r=a},\mathbf{b},\mathbf{c}$. The quantum correlation
probabilities thus become%
\begin{align}
p\left(  a,b\right)   &  =2^{-2(2s-1)}\sin2\xi\cos\left[  2s\left(  \phi
_{a}\mp\phi_{b}\right)  +2\eta\right]  ,\nonumber\\
p\left(  a,c\right)   &  =2^{-2(2s-1)}\sin2\xi\cos\left[  2s\left(  \phi
_{a}\mp\phi_{c}\right)  +2\eta\right]  ,\nonumber\\
p\left(  b,c\right)   &  =2^{-2(2s-1)}\sin2\xi\cos\left[  2s\left(  \phi
_{b}\mp\phi_{c}\right)  +2\eta\right]  , \label{qcr}%
\end{align}
respectively for the antiparallel and parallel spin-polarizations. The front
number-factor $2^{-2(2s-1)}$ would lead to the correlation probability
negligibly small with large spin $s$. This is easy to understand that we
measure only in the four SCSs, while entire dimension of Hilbert space is
$(2s+1)^{2}$. We may consider the relative or scaled correlation probability%
\begin{equation}
p_{rl}\left(  a,b\right)  =\frac{p\left(  a,b\right)  }{N}, \label{sc}%
\end{equation}
where%
\[
N=%
{\displaystyle\sum\limits_{i=1}^{4}}
|\langle i|\psi\rangle|^{2}=%
{\displaystyle\sum\limits_{i=1}^{4}}
\rho_{ii}.
\]
is the total probability of entangled state $|\psi\rangle$ in the four
measuring base-vectors of SCS given by Eq.(\ref{base}).

Using the matrix elements in Eqs.(\ref{nlc11},\ref{el},\ref{nl})
with\ $\theta_{a}=\theta_{b}=\theta_{c}=\pi/2$, the total probability gives
rise to just the same number factor%
\begin{equation}
N=2^{-2(2s-1)}, \label{num}%
\end{equation}
which is $1$ for $s=1/2$. In the following the scaled correlation
probabilities of Eq.(\ref{sc}) is adopted without the subscript "$rl$" for the
sake of simplicity. The scaled quantity of correlation probability difference
Eq.(\ref{bd}) becomes%
\begin{equation}
p_{s}=\sin\left[  2s\left(  \phi_{a}\mp\phi_{b}\right)  \right]  \sin\left[
2s\left(  \phi_{a}\mp\phi_{c}\right)  \right]  -\left\vert \sin\left[
2s\left(  \phi_{b}\mp\phi_{c}\right)  \right]  \right\vert , \label{max}%
\end{equation}
with the polar angles of three measuring directions $\ \theta_{a}=\theta
_{b}=\theta_{c}=\pi/2$ and entangled-state parameters $\xi=\eta=(\pi
/4)\operatorname{mod}2\pi$. From the Eq.(\ref{max}) we can determine the
maximum violation value%
\begin{equation}
p_{s}^{\max}=1 \label{max1}%
\end{equation}
in the case of azimuthal measuring-angles $\phi_{b}=\phi_{c}=0$, and $\phi
_{a}=\pi/2$ or $3\pi/2$. Namely $\mathbf{a}$ is perpendicular to the two
colinear directions $\mathbf{b}$, $\mathbf{c}$. The entangled states to
generate the maximum violation of UBI are respectively%
\[
|\psi\rangle=\frac{1}{\sqrt{2}}\left(  e^{i\frac{\pi}{4}}|+s,-s\rangle
+e^{-i\frac{\pi}{4}}|-s,+s\rangle\right)
\]
and%
\[
|\psi\rangle=\frac{1}{\sqrt{2}}\left(  e^{i\frac{\pi}{4}}|+s,+s\rangle
+e^{-i\frac{\pi}{4}}|-s,-s\rangle\right)  ,
\]
for antiparallel and parallel spin-polarizations in consistence with the
spin-$1/2$ states Eqs.(\ref{ap},\ref{p}).

\section{6. Conclusion}

We reformulate the BI and its violation in a unified formalism by means of the
quantum probability statistics, in which the measuring outcome
correlation-probability is divided into the local and non-local parts. The BI
is derived from the local correlation, while the nonlocal one gives rise to
its violation. Based on the two-spin model we conclude that the violation of
BI depends not only on the specific entangled states but also on the
measurements. For the entangled cat-states the BI is not violated at all
except the spin-$1/2$ case, if the measurements are performed over entire
Hilbert space. The non-local correlation between two components of entangled
state is canceled out completely by the quantum statistical average. On other
hand, when the measurements are restricted in the subspace of SCSs, namely
only the maximum spin values $\pm s$ are taken into account, the UBI proposed
in the present paper is only violated by the entangled states of half-integer
spin but not the integer spin. This spin parity phenomenon is seen to be a
direct result of Berry phase between the SCSs of north- and south- pole
gauges. It is a common belief that there exist two types of non-locality. One
is the phase effect of quantum states such as Aharonov-Bohm and Berry phases.
The other is the violation of BI by entangled states. It is a long standing
problem to find the relation between them. The present paper provides an
example to relate the violation of BI with the geometric phase. The UBI,
$p_{s}^{lc}\leq0$, derived from local-realistic model is suitable to detect
the non-locality of arbitrary entangled cat-states of spin-$s$ for both
antiparallel and parallel polarizations. A maximum violation bound is found as
$p_{s}^{\max}=1$ for the half-integer spin (including the spin-$1/2$) entangled-states.

\section{Acknowledgements}

This work was supported in part by National Natural Science Foundation of
China, under Grants No. 11275118, U1330201.

\section{Appendix}

\subsection{\emph{Following Bell the proof of UBI is trivial from the
classical statistics.}}

The two-spin measuring outcomes (normalized) along the direction $\mathbf{r}$
are denoted respectively by $A(\mathbf{r,}\lambda\mathbf{)}=\pm1$
and\ $B\left(  \mathbf{r},\lambda\right)  =\pm1$ with $\mathbf{r=}$
$\mathbf{a}$, $\mathbf{b}$, $\mathbf{c}$. The (normalized) measuring outcome
correlation for two spins respectively in directions $\mathbf{a}$ and
$\mathbf{b}$ is evaluated with the classical probability statistics under the
local realistic model
\[
p_{lc}\left(  a,b\right)  =\int\rho\left(  \lambda\right)  A\left(
a,\lambda\right)  B\left(  b,\lambda\right)  d\lambda,
\]
in which $\rho\left(  \lambda\right)  $ is the probability density
distribution of parameter $\lambda$. The product of two correlations is
\begin{align*}
&  p_{lc}\left(  a,b\right)  p_{lc}\left(  a,c\right)  \\
&  =\int\int\rho\left(  \lambda\right)  \rho\left(  \lambda^{\prime}\right)
A\left(  a,\lambda\right)  B\left(  b,\lambda\right)  A\left(  a,\lambda
^{\prime}\right)  B\left(  c,\lambda^{\prime}\right)  d\lambda d\lambda
^{\prime}\\
&  \leq\int\rho\left(  \lambda\right)  A^{2}\left(  a,\lambda\right)  B\left(
b,\lambda\right)  B\left(  c,\lambda\right)  d\lambda\\
&  =\int\rho\left(  \lambda\right)  B\left(  b,\lambda\right)  B\left(
c,\lambda\right)  d\lambda
\end{align*}
Using the hidden variable assumption that $B\left(  \mathbf{r},\lambda\right)
=\mathbf{\mp}A\left(  \mathbf{r,}\lambda\right)  $ for the entangled states
respectively with antiparallel and parallel spin-polarizations, we have
\begin{align*}
\int\rho\left(  \lambda\right)  B\left(  b,\lambda\right)  B\left(
c,\lambda\right)  d\lambda &  =\mp\int\rho\left(  \lambda\right)  A\left(
b,\lambda\right)  B\left(  c,\lambda\right)  d\lambda\\
&  \leq\left\vert p_{lc}\left(  b,c\right)  \right\vert
\end{align*}
The UBI%
\[
p_{lc}\left(  a,b\right)  p_{lc}\left(  a,c\right)  \leq\left\vert
p_{lc}\left(  b,c\right)  \right\vert
\]
is suitable to both parallel $\left(  B\left(  \mathbf{b},\lambda\right)
=A\left(  \mathbf{b},\lambda\right)  \right)  $ and antiparallel $\left(
B\left(  \mathbf{b},\lambda\right)  =-A\left(  \mathbf{b},\lambda\right)
\right)  $ polarizations with the total measurement probability $\int%
\rho\left(  \lambda\right)  d\lambda\leq1$, in which the equal sign is valid
only for the spin-$1/2$ states.

\end{document}